\documentclass[aps,prl,floats,twocolumn,superscriptaddress]{revtex4}
\usepackage{amsfonts}
\usepackage{color}
\usepackage{enumerate}
\usepackage{bm}
\usepackage{ dsfont }
\newcommand{\beq}{\begin{equation}}
\newcommand{\eeq}{\end{equation}}
\newcommand{\beqa}{\begin{eqnarray}}
\newcommand{\eeqa}{\end{eqnarray}}
\usepackage{graphicx,amssymb,amsmath,subfigure}
\newcommand{\ket}[1]{\left|#1\right>}
\newcommand{\bra}[1]{\left<#1\right|}

\newcommand{\abs}[1]{\left|#1\right|}

\begin{document}

\title{Linear magnetochiral effect in Weyl semimetals}

\author{Alberto Cortijo}
\email{alberto.cortijo@csic.es}
\affiliation{Instituto de Ciencia de Materiales de Madrid,
CSIC, Cantoblanco; 28049 Madrid, Spain.}

\begin{abstract}
We suggest the possibility of a linear magnetochiral effect in time reversal breaking Weyl semimetals. The magnetochiral effect consists in a simultaneous linear dependence of the magnetotransport coefficients with the magnetic field and a momentum vector. This simultaneous dependence is allowed by the Onsager reciprocity relations, being the separation vector between the Weyl nodes the vector that plays such role. This linear magnetochiral effect constitutes a new transport effect associated to the topological structures linked to time reversal breaking Weyl semimetals.
\end{abstract}
\maketitle
\emph{Introduction.-} In Condensed Matter Physics, a Weyl semimetal (WSM) is an electronic system where the two bands closest to the Fermi level touch each other at a discrete set of points in momentum space (due to the lack of time reversal or inversion symmetries) and whose low energy band structure around these points (Weyl nodes) is described by the so called Weyl Hamiltonian\cite{WTV11},
\beq
H_{W}=\sum_{s=\pm}s v\bm{\sigma}\cdot(\bm{k}-s\bm{b}),\label{HamWeyl}
\eeq
where $2\bm{b}$ is the distance between the Weyl nodes. These systems have attracted much attention nowadays because they are considered as a realization of chiral anomaly-related physics in the Solid State\cite{ZB12}. In short, the chiral anomaly is the non invariance of the Hamiltonian (\ref{HamWeyl}) under chiral phase transformations (i.e. phase transformations with opposite sign for each Weyl node): Since $\bm{b}$ is a constant vector, one can change the phase of the electronic wave function to eliminate it from the Hamiltonian\cite{F85}. The chiral anomaly tells us that this elimination is not totally complete, and the price to pay is the appearance of an anomalous Hall current in the electromagnetic response, $\bm{J}_{H}=\sigma_{H}\bm{b}\times\bm{E}$\cite{WTV11,XWH11}. Other manifestation of the chiral anomaly is the so-called chiral magnetic effect (CME), where in the simultaneous presence of a magnetic and electric fields, a current appears pointing along the magnetic field, $\bm{J}\sim\mu_5\bm{B}$, where $\mu_5$ is the imbalance between the chemical potentials for each Weyl node\cite{FKW08}. Remarkably, this effect is not related to the separation between nodes. The topological character of the CME is the concept behind of the reported longitudinal negative magnetoresistance (LMR) predicted to occur in WSM and measured in several materials candidate to be WSM\cite{KKW13,SS13,HZL15,LKZ16,ASW16,LHL16}. There, a positive contribution to the longitudinal magnetoconductivity (LMC) that scales with $B^2$ is linked to the topological structures associated to the Hamiltonian (\ref{HamWeyl}), as the Berry curvature $\bm{\Omega}(\bm{k})$, and the orbital magnetic moment $\bm{m}(\bm{k})$\cite{XCN10}.
It is stated that in order to observe  effects (others than the $B^2$ contribution described above) associated to the chiral nature of WSMs, or more generically, in chiral conductors, it is necessary to go to the non-linear regime\cite{RF01,MN16}. This is easy to understand: when homogeneous fields are considered, and in absence of any other scale in the problem, the Onsager reciprocity relations forbids any term in the longitudinal conductivity linear in the magnetic field: $\sigma_{33}(\bm{B})=\sigma_{33}(-\bm{B})$ unless the change of sign of $\bm{B}$ is compensated with any other change of sign in the parameters of the problem. In the non-linear case, the injected current itself gives such compensating sign: $\sigma_{33}(\bm{B},\bm{I})=\sigma_{33}(-\bm{B},-\bm{I})$. One then might ask if the vector $\bm{b}$ could provide such compensating sign as well. As we mentioned, this is not allowed by (\ref{HamWeyl}) in virtue of the chiral anomaly.

Besides, there are important theoretical and experimental considerations that call for the need of looking beyond the model (\ref{HamWeyl}). It has been theoretically stated that the CME appears in topologically non-trivial systems that do not support Weyl nodes and are not described by (\ref{HamWeyl})\cite{CY15}. Also, it has been stated that a vector-like coupling between electrons and phonons generically appears in WSM. In order to obtain such coupling, one needs to consider more generic models\cite{CFL15}. Finally, a non-trivial Hall viscosity  appears in the vicinity of a quantum phase transition in WSM when the two Weyl nodes merge\cite{LLS16}. From the experimental side, some materials, like the TaP show up  several Weyl node pairs, being the Fermi level well above the Van Hove (VH) energy (where the two Weyl cones merge into a simply connected dispersion relation) for some of them, and there is a large negative LMR\cite{ASW16}. This situation cannot be considered simply with the Hamiltonian (\ref{HamWeyl}). 

Motivated by these considerations, we will show that a linear magnetochiral effect, that is, a linear behavior of the LMC with the magnetic field $\bm{B}$ appears in simple generalizations of the model (\ref{HamWeyl}) for $\mathcal{T}-$breaking WSMs, implying more consequences of the chirality of WSM than expected \emph{in the linear regime with $\bm{E}$}. Also, we will compute the LMC for Fermi levels above the VH energy, obtaining a non trivial LMC related to the topology of the band structure, and appears in $\mathcal{T}-$preserving WSMs as well. We will show that this linear magnetochiral effect occurs also in the transverse magnetoconductivity (TMC), a transport coefficient that is not related to the anomaly (since $\bm{E}\cdot\bm{B}=0$) but nevertheless related to the topological properties of the system.
\begin{figure*}
\begin{minipage}{.49\linewidth}
(a)
\includegraphics[scale=0.25]{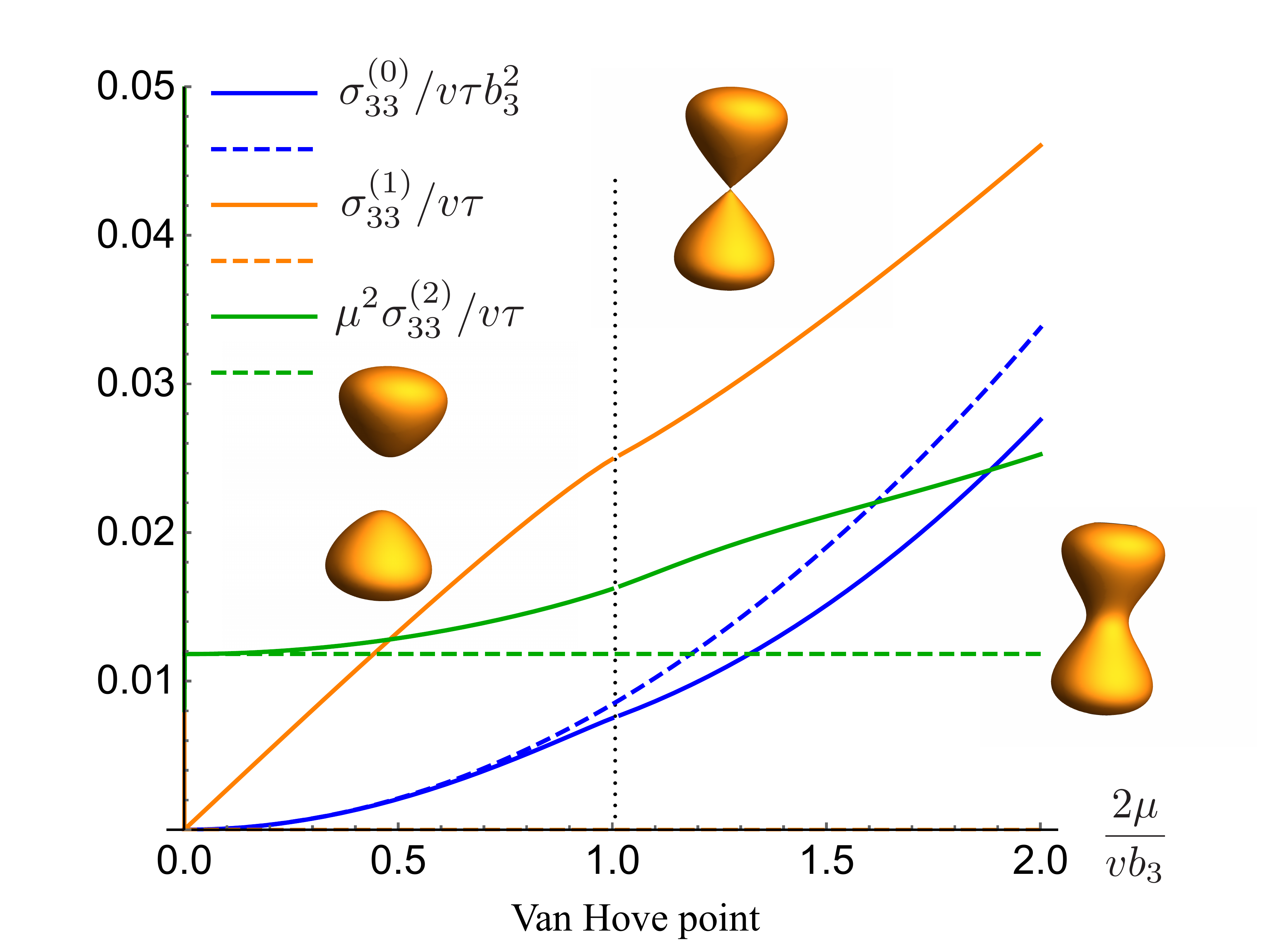}
\end{minipage}
\begin{minipage}{.48\linewidth}
(b)
\includegraphics[scale=0.25]{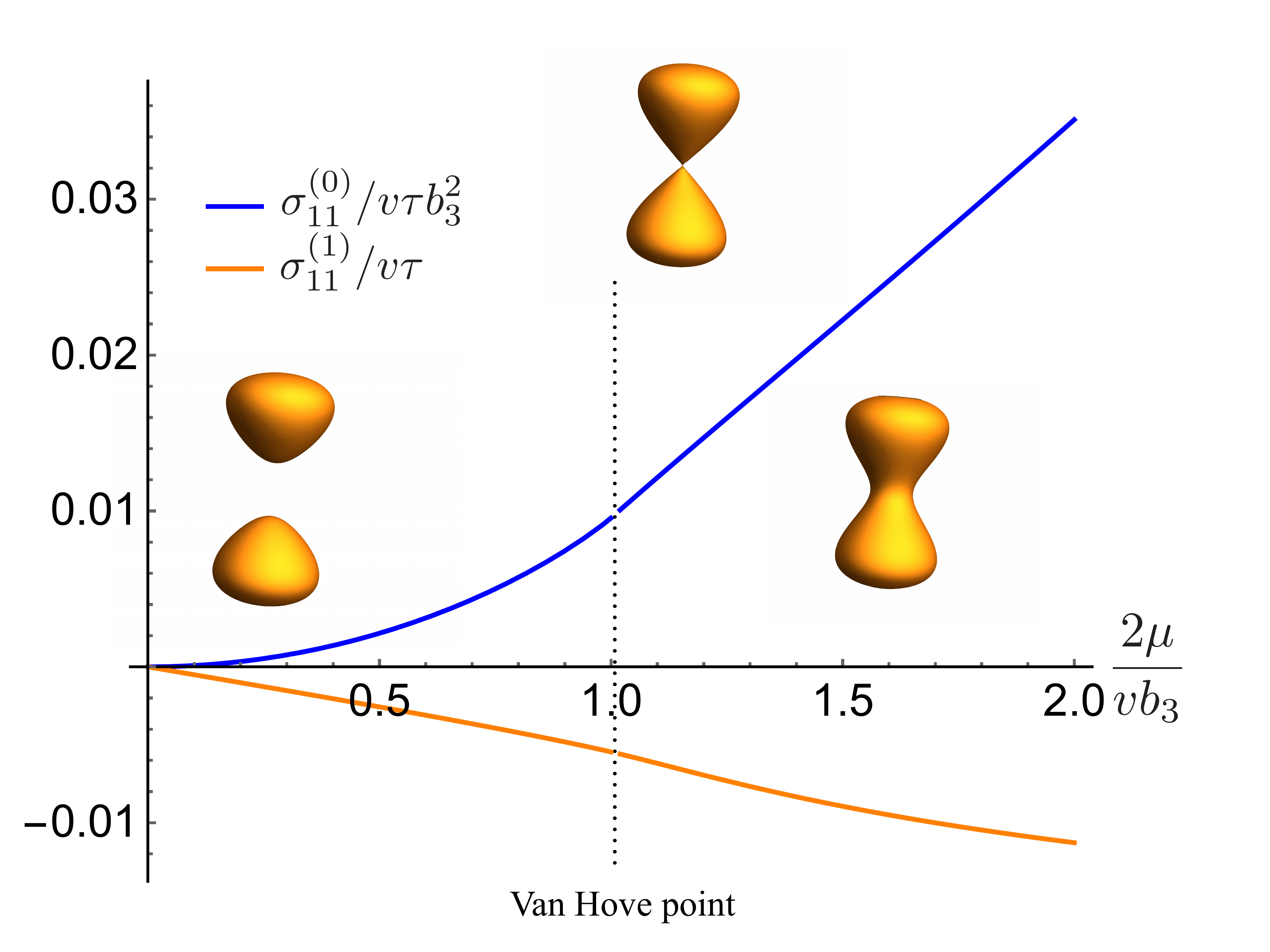}
\end{minipage}
\caption{(Color online)(a) Behavior of the conductivities $\sigma^{(0)}_{33}$, $\sigma^{(1)}_{33}$, and $\mu^2\sigma^{(2)}_{33}$ as a function of the Fermi level $\mu$ (in units of the conductance quantum $e^2/\hbar$). $\sigma^{(2)}_{33}$  appears to be multiplied by $\mu^2$ for clarity. The dashed lines correspond to the conductivities for the model of two unbounded Weyl nodes. Typical shapes of the Fermi surface in each region are plotted. (b) Transverse magnetoconductivity terms as a function of $2\mu/vb_3$. As it happens with the $LMC$, the contribution appears even when there is no well defined chirality ($\mu> \frac{v b_3}{2}$). In both cases we have used $l_{B}=1$ (see the main text) for convenience.}
\label{Figs}
\end{figure*}
In what follows, we will set $\hbar=c=1$ recovering them when the final expressions for the conductivities are presented.

\emph{The model.-} We will employ the following minimal Hamiltonian that generalizes (\ref{HamWeyl}) including band bending effects\cite{LZS15,CY15b,SHS16}:
\beq
H(\bm{k})=v\bm{\sigma}_{\perp}\cdot\bm{k}_{\perp}+\sigma_3 \left(m-\beta k^2_{3}\right),\label{Ham0}
\eeq
with $\bm{\sigma}_{\perp}=(\sigma_1,\sigma_2)$, and $\bm{k}_{\perp}=(k_1,k_2)$. When $m\cdot\beta>0$, this model describes a pair of massless fermions with opposite chirality around the points $\bm{k}^{*}_{\pm}=(\bm{0}_{\perp},\pm b_3)$ with $b_3=\sqrt{m/\beta}$. Close to $\bm{k}^{*}_{\pm}$ and at linear order in momentum, the model (\ref{Ham0}) reduces to (\ref{HamWeyl}) with $v_3=2\sqrt{m\beta}$. We have set $v_3=v$ to make the Weyl fermions isotropic. At energies $\pm\abs{m}$, the model displays Van Hove (VH) singularities, and beyond these points, the notion of chiral fermions is lost.

In order to compute the magnetochiral contribution to the conductivity in the presence of weak electric and magnetic fields, we will resort on the semiclassical kinetic approach\cite{XCN10}.
We will be interested in the situation with external \emph{homogeneous and static} electric and magnetic fields $\bm{E}$ and $\bm{B}$. The program consists in solving the Boltzmann equation for the non-equilibrium distribution function $f^s(\bm{k})$ of the band $s$  in the static and homogeneous case
\beq
\dot{\bm{k}}\cdot\frac{\partial f^s(\bm{k})}{\partial \bm{k}}=-\frac{1}{\tau}\left(f^s(\bm{k})-f^s_{0}(\bm{k})\right),\label{boltzmann1}
\eeq
in the relaxation time approximation, being $f^s_0(\bm{k})=f^s_0(\epsilon^s_0(\bm{k}))$ the Fermi-Dirac distribution function in \emph{absence} of any external electromagnetic field.
To solve (\ref{boltzmann1}), we need the semiclassical equation of motion (SEoM) for the momentum $\bm{k}$ and positions $\bm{x}^s$ of the corresponding Bloch wavepackets for each band $s$: $\dot{\bm{k}}=e\bm{E}+e\dot{\bm{x}}^s\times\bm{B}$, and $\dot{\bm{x}}^s=\bm{v}^s+\dot{\bm{k}}\times\bm{\Omega}^s(\bm{k})$,
where $\bm{v}^s=\frac{\partial \epsilon^s(\bm{k})}{\partial \bm{k}}$ is the group velocity of the dispersion relation $\epsilon^s(\bm{k})$ of the band $s$, modified by the presence of the orbital magnetic moment $\bm{m}^s(\bm{k})$: $\epsilon^s(\bm{k})=\epsilon^s_0(\bm{k})-e\bm{m}^s(\bm{k})\cdot\bm{B}$\cite{XCN10,PKP15,MP15,ZMS16}, and $\bm{\Omega}^s(\bm{k})$ is the Berry curvature associated to the band $s$.

These two equations can be decoupled to get\cite{DHH06}:
\begin{subequations}
\beq
D \dot{\bm{k}}=e\bm{E}+e \bm{v}^s\times \bm{B}+e^2(\bm{E}\cdot \bm{B})\bm{\Omega}^s,\label{forceeq}
\eeq
\beq
D\dot{\bm{x}}^s=\bm{v}^s+e\bm{E}\times\bm{\Omega}^s+e(\bm{v}^s\cdot\bm{\Omega}^s)\bm{B}.\label{velocityeq}
\eeq
\end{subequations} 
with $D=1+e\bm{\Omega}^s\cdot \bm{B}$.


We will linearize the Boltzmann equation (\ref{boltzmann1}) by assuming that $f^s(\bm{k})=f^s_0(\epsilon^s_0(\bm{k}))+f^s_1(\bm{k})$ and keeping first order term $f^s_1(\bm{k})$:

\beq
\frac{D}{\tau}f^s_1(\bm{k})+D\dot{\bm{k}}\cdot \frac{\partial f^s_1(\bm{k})}{\partial \bm{k}}=-D\dot{\bm{k}}\cdot \bm{v}^s_{0} \frac{\partial f^s_0}{\partial \epsilon},\label{linboltzmann}
\eeq
where we have used the fact that $\partial_{\bm{k}} f^s_0=\partial_{\bm{k}} \epsilon^s_0\partial_{\varepsilon} f^s_0=\bm{v}^s_{0}\cdot\partial_{\varepsilon} f_0$.
It is worth to mention that it is often found in the literature\cite{MP15,ZMS16} a different parametrization of the non-equilibrium part $f^s_1(\bm{k})$, because the equilibrium distribution function is defined using the dispersion relation $\epsilon^s(\bm{k})$ modified by the magnetic moment: $f^s(\epsilon^s(\bm{k}))$ instead of $f^s_0(\epsilon^s_0(\bm{k}))$. Since we are computing the magnetoconductivity as a power series of the magnetic field $\bm{B}$, this difference is unessential and one gets the same result at the end of the computations.

The solution of (\ref{linboltzmann}) will be used to compute the non-equilibrium \emph{quasiparticle} current density
\beq
\bm{J}^s=e\int \frac{d^3\bm{k}}{(2\pi)^3}f^s_1(\bm{k})D\dot{\bm{x}}^s.\label{current}
\eeq
Since we are working in the homogeneous limit, there are no contributions others than (\ref{current}) (like the magnetization current) to the total current.

\emph{Longitudinal magnetoconductivity.-}We will employ the method of Jones and Zener\cite{JZ34,PM10} to solve (\ref{linboltzmann}).
In the case of the longitudinal magnetoconductivity we will consider parallel magnetic and electric fields pointing the third spatial direction: $\bm{E}=E_3\hat{\bm{z}}$, and $\bm{B}=B_3\hat{\bm{z}}$.
Also, for definitiveness, we will consider that the Fermi level crosses the conduction band, so, from now on we will drop the band index $s$. To lowest order in the external electric field $\bm{E}$, the solution reads\footnote{See the Supplemental Material for (i) details of the model, (ii) the solution of the Boltzmann equation, (iii) the computation of each contribution to the conductivities, and (iv) the proof that the linear magnetochiral term enters as the scalar product between $\bm{b}$ and the magnetic field.}
\beq
f_1(\bm{k})=-\frac{\tau}{D}\frac{\partial f_0}{\partial \epsilon} \left( v_{3}(\bm{k})+eB_3 \Omega_a(\bm{k}) v_{a}(\bm{k})\right)eE_3.\label{finalboltzmann2}
\eeq

Inserting (\ref{finalboltzmann2}) and (\ref{velocityeq}) into (\ref{current}), employing the definition of $v_i(\bm{k})$, and expanding the integrand in powers of $eB_3$ up to quadratic order in $eB_3$, we obtain the following expressions for the terms of the magnetoconductivity $\sigma_{33}[B_3]\simeq \sigma^{(0)}_{33}+\sigma^{(1)}_{33}[B_3]+\sigma^{(2)}_{33}[B_3^2]+...$: 
\begin{widetext}
\begin{subequations}
\beq
\sigma^{(0)}_{33}=-e^2\tau\int \frac{d^3\bm{k}}{(2\pi)^3}\frac{\partial f_0}{\partial \epsilon}\left(v^0_{3}\right)^2,\label{contri0}
\eeq
\beq
\sigma^{(1)}_{33}=-e^3B_3\tau\int\frac{d^3\bm{k}}{(2\pi)^3}\frac{\partial f_0}{\partial \epsilon}\left(2\Omega_a v^0_{a}-\partial_3 m_3-v^0_{3}\Omega_3\right)v^0_{3},\label{contri1}
\eeq
\beq
\sigma^{(2)}_{33}=-e^4B^2_3\tau\int \frac{d^3\bm{k}}{(2\pi)^3}\frac{\partial f_0}{\partial \epsilon}\left((\Omega_a v^0_{a}-\Omega_3v^0_{3})^2-(\Omega_av^0_{a}-v^0_{3}\Omega_3 )\partial_3 m_3-v^0_{3}\Omega_a\partial_a m_3\right),\label{contri2}
\eeq
\label{conductivities}
\end{subequations}
\end{widetext}
where it is understood that $\partial_a m_3=\partial m_3/\partial k_a$.

We will work at zero temperature ($T=0$), so the  derivative of the equilibrium distribution function $f_{0}$ will be strongly peaked at the Fermi level $\mu$: $\partial f_0/\partial \epsilon\simeq-\delta\left(\mu-\epsilon_0(\bm{k})\right)$.

In principle, fully analytic expressions for the conductivity $\sigma_{33}$ can be found as a function of the Fermi level $\mu$ and $b_3$, thought they do not provide too much physical insight. What it is interesting is to know their behavior when the Fermi level lies well below the VH energy $\epsilon_{VH}$ ($\mu\ll vb_3$), or when $\mu\gg vb_3$,  corresponding to the situation when the Fermi level is well above $\epsilon_{VH}$ or the Weyl node separation is too small, situations where the notion of chirality is lost\cite{ASW16}. 

The zeroth order of the conductivity, $\sigma^{(0)}_{33}$ is standard. For $\mu\ll vb_3$, we obtain (recovering $\hbar$ and $c$)
\beq
\sigma^{(0)}_{33}\simeq\frac{e^2\tau v}{\pi^2\hbar }\frac{\mu^2}{v^2}\left(\frac{1}{3}-\frac{1}{10}\frac{\mu^2}{v^2b^2_3}\right).\label{sigma0low}
\eeq
The first term in the parenthesis can be found in the literature\cite{SS13}, when the usual model for a pair of \emph{unbounded}, linearly dispersing chiral fermions is considered. The second term in the parenthesis is a correction due to the modification of the Hamiltonian (\ref{HamWeyl}).
In the oposite limit, $\mu\gg vb_3$, we obtain
\beq
\sigma^{(0)}_{33}\simeq \frac{e^2\tau v}{7\pi^2\hbar }\frac{\mu^2}{v^2}\left(\frac{2\mu}{v b_3}\right)^{\frac{1}{2}}.\label{sigma0high}
\eeq

The interesting situation comes with $\sigma^{(1)}_{33}$. For $\mu\ll vb_3$, one gets
\beq
\sigma^{(1)}_{33}=\frac{8e^3\tau v}{15\pi^2\hbar c}\frac{\mu}{v b_3}B_3.\label{sigma1low}
\eeq

Written in this way, eq.(\ref{sigma1low}) does not show all its physical content. Remembering that the positions of the Weyl nodes $\bm{b}$ are actually vectors in the momentum space, we can rewrite $\sigma^{(1)}_{33}$ as (see Supplemental Material)
\beq
\sigma^{(1)}_{33}\simeq \frac{8e^3\tau v}{15 \pi^2\hbar c}\frac{\mu}{v|\bm{b}|^2}\bm{b}\cdot\bm{B}.\label{sigma1low2}
\eeq
When the model of a pair of linearly dispersing unbounded Weyl fermions is considered, the contribution $\sigma^{(1)}_{33}$ is directly zero. As we mentioned above, the chiral anomaly forces the anomalous Hall conductivity to be the only term proportional to the vector $\bm{b}$.

Since, as we said, $\bm{b}$ a vector in momentum space, it is odd under time reversal symmetry, $\bm{b}\rightarrow-\bm{b}$, as it is $\bm{B}$, so $\sigma^{(1)}_{33}(\bm{b}\cdot\bm{B})=\sigma^{(1)}_{33}(-\bm{b}\cdot(-\bm{B}))$, as it is dictated by the Onsager reciprocity principle. This term is the electrical magnetochiral term in the magnetoconductivity\cite{RF01}. Quite remarkably, looking at the expression (\ref{contri1}), this effect stems from the presence of a non vanishing Berry curvature and the orbital magnetic moment. 
In the opposite limit $\mu\gg vb_3$,
\beq
\sigma^{(1)}_{33}\simeq\frac{15e^3\tau v}{77\pi^2\hbar c}\frac{\mu}{v b_3}\left(\frac{2\mu}{v b_3}\right)^{\frac{1}{2}}B_3.\label{sigma1high}
\eeq
Finally, the term $\sigma^{(2)}_{33}$ reads, for $\mu\ll vb_3$,
\beq
\sigma^{(2)}_{33}\simeq\frac{e^4\tau v}{2\pi^2\hbar c^2}\left(\frac{7}{30}\frac{v^2}{\mu^2}+\frac{47}{140}\frac{1}{b^2_3}\right)B^2_3,\label{sigma2low}
\eeq
Apart from the numerical factor, the first term in (\ref{sigma2low}) is the term associated to the chiral anomaly in the case of a pair of linearly dispersing Weyl fermions\cite{SS13}. The apparent discrepancy comes from the fact that here we have taken into account explicitly the effect of the orbital magnetic moment. Interestingly, the next term in (\ref{sigma2low}) is not depending on the Fermi level but depends only on the (squared) length of vector $\bm{b}$.

In the opposite regime, well above the VH point (when $\mu\gg vb_3$) we obtain
\beq
\sigma^{(2)}_{33}\simeq \frac{16}{165}\frac{e^4\tau v}{\pi^2\hbar c^2}\frac{1}{b^2_3}\left(\frac{2\mu}{v b_3}\right)^{\frac{1}{2}}B^2_3.\label{sigma2high}
\eeq

The longitudinal magnetoconductivity ($LMC$) defined as $LMC=\frac{\sigma_{33}[e\mathbf{B}]-\sigma_{33}[0]}{\sigma_{33}[0]}$ is, in the limit $\mu\ll vb_3$,
\beq
LMC\simeq \frac{e}{c}\frac{16}{5}\frac{v}{\mu |\bm{b}|^2}\bm{b}\cdot\bm{B}.\label{LLMClow}
\eeq

Let us discuss the constraints that time reversal symmetry $\mathcal{T}$ imposes on $\sigma^{(1)}_{33}$. It is well known that the model (\ref{Ham0}) breaks $\mathcal{T}$, having associated a topological Hall conductivity proportional to $\bm{b}$. It is easy to see that the time reversal invariant partner of the Hamiltonian (\ref{Ham0}) can be constructed simply by replacing $d_{3}(\bm{k})$ by $-d_{3}(\bm{k})$ and a trivial unitary transformation. We can repeat all the steps and compute $\sigma_{33}$ for the time reversal partner of (\ref{Ham0}), and obtain $\sigma^{(1)\mathcal{T}}_{33}=-\sigma^{(1)}_{33}$, so, for time reversal invariant Weyl semimetals, the magnetochiral effect proposed here vanishes, while $\sigma^{(0)\mathcal{T}}_{33}=\sigma^{(0)}_{33}$, and $\sigma^{(2)\mathcal{T}}_{33}=\sigma^{(2)}_{33}$.

There are several ways of braking $\mathcal{T}$. For instance, the simplest scenario consists in two pairs of Weyl nodes related by $\mathcal{T}$, where the separation between Weyl nodes is different for each Kramer partner (represented by $\uparrow,\downarrow$): $|\bm{b}|_{\uparrow}>|\bm{b}|_{\downarrow}$, denoting  $\mu_{\uparrow}$ and $\mu_{\downarrow}$ as the respective Fermi levels, the result is simply,
\beq
\sigma^{(1)}_{33}\simeq \frac{8e^3\tau v}{15 \pi^2\hbar c}\left(\frac{\mu_{\uparrow}}{|\bm{b}_{\uparrow}|^2}\bm{b}_{\uparrow}-\frac{\mu_{\downarrow}}{|\bm{b}_{\downarrow}|^2}\bm{b}_{\downarrow}\right)\cdot\bm{B}.
\eeq
This expression paves the way for computing more complex configurations.

\emph{Transverse magnetoconductivity.-} In the same manner we computed the linear $LMC$ ($\bm{E}||\bm{B}$), we can compute the linear transverse magnetoconductivity ($LTMC$) for electric fields $\bm{E}\perp\bm{B}$. Without loss of generality, we will focus in an electric field $\bm{E}=E_1\hat{\bm{x}}$ keeping $\bm{B}=b_3\hat{\bm{z}}$. Strictly speaking, it is clear that no term in the $TMC$ can be associated to the chiral anomaly ($\bm{E}\cdot\bm{B}=0$) with this electromagnetic configuration, and yet, there are terms in the $TMC$ coming from the topological structures associated to the bandstructure, $\bm{\Omega}(\bm{k})$ and $\bm{m}(\bm{k})$.  It can be shown  that the transverse conductivity $\sigma_{11}[B_3]\simeq\sigma^{(0)}_{11}+\sigma^{(1)}_{11}[eB_3]$ is (retaining the non-vanishing terms after momentum integration)
\begin{subequations}
\beqa
\sigma^{(0)}_{11}&=&-e^2\tau \int \frac{d^3\bm{k}}{(2\pi)^3}\frac{\partial f_0}{\partial \epsilon}\left(v^0_1\right)^2\simeq\\\nonumber
&\simeq&\frac{1}{3}\frac{e^2\tau v}{\pi^2\hbar} \frac{\mu^2}{v^2}.
\eeqa
\beqa
\sigma^{(1)}_{11}&=&e^3\tau B_3 \int \frac{d^3\bm{k}}{(2\pi)^3}\frac{\partial f_0}{\partial \epsilon}\left(\partial_1 m_3+\Omega_3 v^{0}_1\right)v^{0}_1\simeq\\\nonumber
&\simeq&-\frac{1}{10}\frac{e^3\tau v}{\pi^2\hbar c}\frac{\mu}{vb_3}B_3,\label{sigmaT1}
\eeqa
\end{subequations}
in the limit $\mu\ll vb_3$. With these two expressions, to lowest order in the magnetic field, the linear $TMC$ takes the following form ($\mu\ll vb_3$):
\beq
TMC=\frac{\sigma_{11}[e\bm{B}]-\sigma_{11}[0]}{\sigma_{11}[0]}\simeq -\frac{e}{c}\frac{3}{10}\frac{v}{\mu |\bm{b}|^2}\bm{b}\cdot\bm{B},\label{LTMClow}
\eeq
that is, there is also an electric magnetochiral effect in the $TMC$, but in opposition to the $LMC$, the $TMC$ turns out to be \emph{negative}.

\emph{Conclusions.-} We have described a topologically related magnetochiral effect in WSMs that break $\mathcal{T}$. So far, there are several candidates of $\mathcal{T}-$breaking type-I\cite{LHZ15,CSX16,WVK16} and type-II WSMs\cite{BEG15} so the theory of linear magnetochiral effect presented here can be experimentally tested. We have shown that the band bending terms (beyond the linear in momentum term) are responsible of this effect. This magnetochiral term is topological in origin, as the $B^2$ term in the MC since it is related to the topological properties of the band structure, $\bm{\Omega}(\bm{k})$ and $\bm{m}(\bm{k})$. Besides, the band bending terms are unavoidable in realistic solid state-based WSMs, the magnetochiral effect will be inevitably present in time reversal breaking WSMs.

For $\mathcal{T}-$preserving WSMs, we have shown that the topological $B^2$ contribution to the LMC is still present when the Fermi level $\mu$ is above the VH energy, and there are contributions to this part of the LMC that are independent of $\mu$.
It is important to stress that the magnetochiral effect obtained here appears in the regime of small magnetic fields, that is, the magnetic length $l_{B}\sim(eB)^{-1/2}$ being the largest scale of the problem: $l_{B}\gg v\tau$, $l_{B}\gg1/|\bm{b}|$, and $l_{B}\gg v/\mu\gg1$, limits that define the range of validity of the semiclassical kinetic theory). In the opposite regime, corresponding to the ultraquantum limit, a linear LMC has been theoretically reported\cite{GMS14,GPS15,SHS16}.
While the magnetochiral effect described here is due to the presence of band bending terms in the Hamiltonian (\ref{Ham0}) the reported linear behavior in the ultraquantum limit stems from the chiral anomaly activated in the lowest Landau level and concrete forms of disorder potentials.

In the present paper we have focused on the magnetochiral effect appearing in the LMC and TMC in the static and homogeneous limit. In principle there is no reason to not expect this magnetochiral effect in thermal transport \cite{HKW16} or in optical properties associated to (\ref{Ham0}). We leave such questions for future research.

\emph{Acknowledgements.-} The author acknowledges fruitful discussions with Y. Ferreiros, K. Landsteiner, and M. A. H. Vozmediano. Financial support from the European Union structural funds and the Comunidad de Madrid MAD2D-CM Program (S2013/MIT-3007) and MINECO (Spain) Grant No. FIS2015-73454-JIN is acknowledged.


\newpage
\appendix
\begin{widetext}
\section{SUPPLEMENTAL MATERIAL FOR LINEAR MAGNETOCHIRAL EFFECT IN WEYL SEMIMETALS}
\subsection{The model and computation of the Berry curvature and the orbital magnetic moment}
\subsubsection{The model}
In the main text, we have used the model
\beq
H(\bm{k})=v\sigma_{\perp}\cdot\bm{k}_{\perp}+\sigma_3(m-\beta k^2_3),\label{SHam000}
\eeq
with $m$ and $\beta$ two paramenters that represent an energy scale and a band bending parameter (with dimensions of energy times squared length) respectively. This model has been discussed in the literature in several times (see references 18, 19, and 20 of the main text). As described in the main text, it possesses two Weyl points at momenta $(\bm{0}_{\perp},\pm b_3)$, with $b_3=\sqrt{m/\beta}$. Around these two points, the dispersion relation is linear in momentum, with velocity $v=2\sqrt{m\beta}$. As a simplification, we will assume that the dispersion relation around these two Weyl points is isotropic. This simplification does not alter the results presented in this work. The previous model posesses the same constraints that the linear Weyl Hamiltonian, that is, having the two Weyl nodes at the same energy and exactly at opposite values of momenta in the momentum space. Also the model (\ref{SHam000}) is axisymmetric around the axis defined by $\bm{b}=(\bm{0}_{\perp},b_{3})$.

In all the calculations that follow, it turns out to be more convenient to rewrite this Hamiltonian, not in terms of $m$ and $\beta$ but in terms of $v$ and $b_3$. It is a simple substitution to find that $m=\frac{1}{2}v b_3$ and $\beta=\frac{v}{2b_3}$, so the model can be written as
\beq
H(\bm{k})=v\sigma_{\perp}\cdot\bm{k}_{\perp}+\sigma_3 \frac{vb_3}{2}\left(1-\frac{k^2_3}{b^2_3}\right).\label{SHam00}
\eeq
This is the form we will use in all the calculations in what follows. 

The model (\ref{SHam000}) admits an extra term  of the form of $\beta_{\perp}\bm{k}^2_{\perp}$, with $\beta_{\perp}\cdot m>0$. The presence of this term does not modify neither the position of the Weyl points nor the linear dispersion around them. They do not modify the qualitative results presented in this work, and it only adds extra complexity in the analytical calculations presented below. For these reasons, from now on we will neglect it.

\subsubsection{Berry curvature and magnetic moment}
Let us briefly review the models and concepts that we will use later. We will deal with two main models: a pair of unbounded linearly dispersive Weyl particles and a two-band model where the bands cross each other defining in the low energy regime a pair of Weyl nodes. Both Hamiltonian models fit into the general expression ($s=\pm1$ refers to the valence and conduction bands)
\beq
H=\bm{\sigma}\cdot\bm{d}(\bm{k}),\label{SHam0}
\eeq 
Where $\bm{\sigma}$ are the Pauli matrices and $\bm{d}(\bm{k})$ is a vector depending on the momentum (as usual $\hbar=1$ and $c=1$. The $c$ is easy to restore: each power of magnetic field $\bm{B}$ will have a $c$ dividing). For the Weyl mode, we will use $d^{s}_1(\bm{k})=sv k_1$, $d^s_2(\bm{k})=sv k_2$, and $d^s_3(\bm{k})=svk_3$, being $s=\pm 1$. Each Weyl node has opposite chirality denoted by $s$, so, actually we should write $H=\sum_s H^s$.
For the two band model, we will use $d_1(\bm{k})=vk_1$, $d_2(\bm{k})=vk_2$ and $d_{3}(\bm{k})=\frac{vb_3}{2}(1-\frac{k^2_3}{b^2_3})$. The choice of the parameters is to get the Weyl nodes at $k_c=\pm b_3$ and a dispersion $\pm v\delta k_3$ around each node. In any case, we will work out with the generic model (\ref{SHam0}). Intentionally, the two models have the property of each component of $d_i$ depending only on the corresponding momentum component $k_i$, that is, $\frac{\partial d_i}{\partial k_j}= \delta_{ij}\frac{\partial d_i}{\partial k_i}$.

Explicit expressions for the group velocity $v^0_i=\partial \epsilon_{\bm{k}}/\partial k_i$, the orbital magnetic moment $m_i$ and the Berry curvature $\Omega_i$ will be necessary, so we proceed to compute them:

\beq
H\ket{s}=\bm{\sigma}\cdot\bm{d}(\bm{k})\ket{s}=s\epsilon_{\bm{k}}\ket{s},\qquad \epsilon_{\bm{k}}=|\bm{d}(\bm{k})|.
\eeq
As discussed in the main text, we will place the chemical potential crossing the conduction band so in what follows we will compute everything for the conduction band, $s=+$:

\beq
v^0_{i}=\frac{\partial \epsilon_{\bm{k}}}{\partial k_i}=\frac{\partial d_j}{\partial k_i}\frac{\partial \epsilon_{\bm{k}}}{\partial d_{j}}=\left(\frac{\partial d_i}{\partial k_i}\right)\frac{d_i}{\epsilon_{\bm{k}}}.
\eeq

\beqa
\Omega_i=i\varepsilon_{ijl}\langle \partial_j + | \partial_l + \rangle=i\epsilon_{ijl} \langle \partial_j + |-\rangle \langle -| \partial_l + \rangle.
\eeqa

The part with $\ket{+}\bra{+}$ is identically zero because the contraction of the Levi-Civitta symbol and something symmetric (it is not difficult to see this by noticing that $\langle s^\prime | \partial_i s\rangle=-\langle\partial_i s^\prime|s\rangle$). Using

\beq
\langle -| \partial_l + \rangle=\frac{1}{2\epsilon_{\bm{k}}}\bra{-}\partial_{l}H\ket{+},
\eeq
and
\beq
\partial_l H=\sigma_l \frac{\partial d_l}{\partial k_l},
\eeq
we get
\beqa
\Omega_i=\frac{i}{4\varepsilon^2_{\bm{k}}}\epsilon_{ijl}\frac{\partial d_j}{\partial k_j}\frac{\partial d_l}{\partial k_l}\bra{+}\sigma_{j}\ket{-}\bra{-}\sigma_l \ket{+}=
\frac{i}{4\varepsilon^2_{\bm{k}}}\epsilon_{ijl}\frac{\partial d_j}{\partial k_j}\frac{\partial d_l}{\partial k_l}\bra{+}\sigma_{j}\sigma_l \ket{+}.
\eeqa
Now, the last equality occurs because we are trivially completing the basis ($\epsilon_{ijl}\bra{+}\sigma_j \ket{+}\bra{+}\sigma_{l}\ket{+}=0$ since we are multiplying the Levi-Civitta symbol by a symmetric product of vectors, and $\bra{+}\sigma_i\ket{+}=\frac{d_i}{\varepsilon_{\bm{k}}}$).

 Using $\sigma_j\sigma_l=\delta_{jl}\sigma_0+i\epsilon_{jlr}\sigma_r$, we get
\beqa
\Omega_i=-\frac{1}{2\varepsilon^2_{\bm{k}}}\left(\frac{\partial d_i}{\partial k_i}\right)^2\bra{+}\sigma_i\ket{+}=-\frac{1}{2\varepsilon^3_{\bm{k}}}\left(\frac{\partial d_i}{\partial k_i}\right)^2 d_i.
\eeqa

We can proceed in the same way for the orbital magnetization for the conduction band, defined as
\beq
m_{i}=\frac{i}{2}\epsilon_{ijl}\bra{\partial_j +}H-\epsilon_{\bm{k}}\ket{\partial_l +}. 
\eeq
Using the same tricks as before, one gets
\beq
m_{i}=\frac{1}{2\varepsilon^2_{\bm{k}}}\left(\frac{\partial d_i}{\partial k_i}\right)^2 d_i.
\eeq

We can use these general expressions to obtain the Berry curvature and orbital magnetic moment for the model (\ref{SHam00}). By simple substitution one gets:
\begin{subequations}
\beq
\bm{\Omega}_{\perp}=-\frac{1}{2}\frac{\bm{k}_{\perp}}{(\bm{k}^2_{\perp}+(\frac{b_3}{2}-\frac{k^2_3}{2b_3})^2)^{3/2}},
\eeq
\beq
\Omega_3=-\frac{1}{4b_3}\frac{k^2_3(1-\frac{k^2_3}{b^2_3})}{(\bm{k}^2_{\perp}+(\frac{b_3}{2}-\frac{k^2_3}{2b_3})^2)^{3/2}},
\eeq 
\end{subequations}
and 

\begin{subequations}
\beq
\bm{m}_{\perp}=\frac{1}{2}\frac{v\bm{k}_{\perp}}{(\bm{k}^2_{\perp}+(\frac{b_3}{2}-\frac{k^2_3}{2b_3})^2)},
\eeq
\beq
m_3=-\frac{1}{4b_3}\frac{v k^2_3(1-\frac{k^2_3}{b^2_3})}{(\bm{k}^2_{\perp}+(\frac{b_3}{2}-\frac{k^2_3}{2b_3})^2)}.
\eeq 
\end{subequations}
\subsection{Detailed solution of the Boltzmann equation}
Here I am quoting the semiclassical equations of motion. In principle, the form of the equations of motion does not rely on any particular model Hamiltonian, but the only requirement is that the Berry connection $\mathcal{A}^s_{i}(\bm{k})=i\langle s|\partial_i s\rangle$ is \emph{abelian} ($\Omega_i=\epsilon_{ijl}\partial_j \mathcal{A}_l$):
\begin{subequations}
\beq
\dot{k}_i=e E_i+e\epsilon_{ijl}\dot{x}_j B_l,
\eeq
\beq
\dot{x}_{i}=v_i+\epsilon_{ijl}\dot{k}_j\Omega_l, 
\eeq
\end{subequations}
with $v_i=v^0_i-e\partial_{i}m_a B_a=\partial_i \varepsilon_{\bm{k}}-e\partial_{i}m_a B_a$, as the generalized velocity. Denoting the product $D=1+e\Omega_i\cdot B_i$, the well known solutions are
\begin{subequations}
\beq
D \dot{k}_{i}=eE_i+e\epsilon_{ijl} v_j B_l+e^2(E_a\cdot B_a)\Omega_i,\label{Sforceeq}
\eeq
\beq
D\dot{x}_i=v_i+e\epsilon_{ijl}E_j\Omega_l+e(v_a\cdot\Omega_a)B_i.\label{Svelocityeq}
\eeq
\end{subequations} 
It is interesting to note that in the first equation (\ref{Sforceeq}), representing the classical second Newton law, the term associated to the Berry curvature represents a deviation of the standard Lorentz force, in the sense that the effective electric field experienced by the electronic wave packet is $\mathcal{E}_i=(\delta_{ir}+e\Omega_{i}B_r)E_r$, not pointing exactly along $E_i$.

We will use the \emph{quasiparticle current}, defined as ($(d\bm{k})\equiv\frac{d^3\bm{k}}{(2\pi)^3}$):
\beq
J_{i}=e\int (d\bm{k})f(\bm{k})D\dot{x}_i,\label{Scurrent}
\eeq
with $f(\bm{k})$ is the non-equilibrium distribution function and $D\dot{x}_i$ from (\ref{Svelocityeq}). 

The next step is to solve the Boltzmann equation in the static and homogeneous configuration
\beq
\dot{k}_i\frac{\partial f(\bm{k})}{\partial k_i}=I_{c}[f(\bm{k})],\label{SBoltzmann}
\eeq
with $I_c[f(\bm{k})]$ is the collision integral. We will employ the \emph{relaxation time approximation}:
\beq
I_c[f(\bm{k})]=-\frac{f(\bm{k})-f_{0}(\bm{k})}{\tau_{\bm{k}}}.
\eeq

Here $\tau_{\bm{k}}$ is the relaxation time. As usual, the non-equilibrium distribution function $f(\bm{k})$ will decomposed in the equilibrium distribution function $f_0(\bm{k})$ in absence of any electromagnetic field, and a perturbation $f_1(\bm{k})$, $f(\bm{k})=f_0(\bm{k})+f_1(\bm{k})$. Inserting this in the Boltzmann equation (\ref{SBoltzmann}) together with (\ref{Sforceeq}) we obtain the linearized Boltzmann equation:
\beq
\frac{D}{\tau_{\bm{k}}}f_1+D\dot{k}_i\cdot \frac{\partial f_1}{\partial k_i}=-D\dot{k}_i\cdot v^0_i \cdot\frac{\partial f_0}{\partial \varepsilon_{\bm{k}}},\label{Slinboltzmann}
\eeq
where we have used the fact that $f_0(\bm{k})=f_0(\varepsilon_{\bm{k}})$ so $\frac{\partial f_0}{\partial k_i}=v^0_i\cdot\frac{\partial f_0}{\partial \varepsilon_{\bm{k}}}$. We have also multiplied the whole equation by $D$.

Let us carefully solve equation (\ref{Slinboltzmann}). The product $D\dot{k}_i\cdot v^0_i$ reads:
\beqa
D\dot{k}_i\cdot v^0_i= eE_iv^0_i+e^2(E_a\cdot B_a)\Omega_i v^0_i-e^2\epsilon_{ijl}v^0_i(\partial_j m_a)B_a B_l.
\eeqa
We will choice the electromagnetic configuration $\bm{B}=B_3\hat{z}$, so first, $D=1+e\Omega_3 B_3$, and 
\beqa
D\dot{k}_i\cdot v^0_i=eE_iv^0_i+e^2(E_3 B_3)\Omega_i v^0_i-e^2\epsilon_{ij3}v^0_i(\partial_j m_3)B^2_3.\label{Skdoteq}
\eeqa
Let us work out a little bit more the last term on the RHS. Going to cylindrical coordinates ($k_1=k\cos\theta$, $k_2=k\sin\theta$, and $k_3=k_3$, so $\partial_1=\cos\theta\partial_k-\frac{1}{k}\sin\theta\partial_\theta$, and $\partial_2=\sin\theta\partial_k+\frac{1}{k}\cos\theta\partial_\theta$), for both models, $d_1=vk_1$ and $d_2=vk_2$ so:
\beq
v^0_1\partial_2m_3-v^0_2\partial_1m_3=\frac{v^2}{\varepsilon_{\bm{k}}}\partial_\theta m_3.
\eeq

Now comes an important remark. For much more complicated situations, $d_3$ might be a complex function of the polar angle $\theta$, so $\partial_{\theta}m_3$ can be non-zero. Importantly, in the two models chosen, we will assume that $d_3$ is only function of $k_3$ and not on $\theta$ so the same will happen for $m_3$ and $\varepsilon_{\bm{k}}$. Under such approximation,
\beq
\partial_\theta m_3=\partial_\theta \frac{1}{2\varepsilon^2_{\bm{k}}}\left(\frac{\partial d_3}{\partial k_3}\right)^2 d_3=0,
\eeq
and then
\beq
D\dot{k}_i=E_iv^0_i+e^2(E_3 B_3)\Omega_i \cdot v^0_i.
\eeq
Now we can see that all the terms of the RHS of equation (\ref{Slinboltzmann}) are linear in the electric field $\bm{E}$, so, assuming that we work in the linear response regime to the electric field, we can consider that $f_1\sim O(E_i)$, so, to linear order in $\bm{E}$, the linearized boltzmann equation is 
\beq
\frac{D}{\tau_{\bm{k}}}f_1+eB_3\epsilon_{ij3}v_j\frac{\partial f_1}{\partial k_i}=-\frac{\partial f_0}{\partial \varepsilon_{\bm{k}}}(eE_i v^0_i+e^2E_3B_3\Omega_i\cdot v^0_i).
\eeq
From now on, we will simply denote the transport time as $\tau$.

Now let us work out the second term of the LHS of this equation. Using again polar coordinates, it is easy to see that 
\beq
v^0_2\partial_1-v^0_1\partial_2=-v^2\frac{1}{\varepsilon_{\bm{k}}}\frac{\partial}{\partial \theta},
\eeq
and (remembering that $\partial_\theta m_3=0$)
\beq
\partial_2 m_3\partial_1-\partial_1 m_3\partial_2=-\frac{1}{k}\partial_k m_3\partial_\theta,
\eeq
so
\beqa
\frac{D}{\tau}f_1-eB_3\left(\frac{v^2}{\varepsilon_{\bm{k}}}-eB_3\frac{1}{k}\partial_k m_3\right)\frac{\partial f_1}{\partial \theta}=
\frac{\partial f_0}{\partial \varepsilon_{\bm{k}}}\left(eE_i v^0_i+e^2 E_3B_3 \Omega_i v^0_i\right).
\eeqa
Since neither $\tau$ nor $D$ depend on $\theta$, we can multiply everything by $\tau/D$ and write the former equation in a more compact way:
\beqa
\left(1+eB_3\tau g\frac{\partial}{\partial \theta}\right)f_1=
\frac{\tau}{D}\frac{\partial f_0}{\partial \varepsilon_{\bm{k}}}\left(eE_i v^0_i+e^2 E_3B_3 \Omega_i v^0_i\right),
\eeqa
with $g=\left(\frac{eB_3}{D}\frac{1}{k}\partial_k m_3-\frac{v^2}{D\varepsilon_{\bm{k}}}\right)$. 
Cleaning up a little bit more the notation, let us define the differential operator $\hat{\Theta}=eB_3\tau g\frac{\partial}{\partial \theta}$ and rewrite the Boltzmann equation as
\beqa
\left(1+\hat{\Theta}\right)f_1=-\frac{\tau}{D}\frac{\partial f_0}{\partial \varepsilon_{\bm{k}}}\left(eE_i v^0_i+e^2 E_3B_3 \Omega_i v^0_i\right).
\eeqa
It turns out that the linearized Boltzmann equation is an inhomogeneous first order differential linear equation in the operator $\hat{\Theta}$ and it always has a solution, that can be computed explicitly. However, is way much cheaper to use the Zener-Jones method for solving it by formally writing
\beqa
f_1=-\frac{\tau}{D}\frac{\partial f_0}{\partial \varepsilon_{\bm{k}}}\left(\frac{1}{1+\hat{\Theta}}\right)\left(eE_i v^0_i+e^2 E_3B_3 \Omega_i v^0_i\right),
\eeqa
and defining the inverse operator as an infinite series of operators:
\beq
\left(\frac{1}{1+\hat{\Theta}}\right)=1+\sum^{\infty}_{n=1}(-\hat{\Theta})^n,
\eeq
so
\beqa
f_1=-\frac{\tau}{D}\frac{\partial f_0}{\partial \varepsilon_{\bm{k}}}\left(eE_i v^0_i+e^2 E_3B_3 \Omega_i v^0_i\right)-
\frac{\tau}{D}\frac{\partial f_0}{\partial \varepsilon_{\bm{k}}}\sum^{\infty}_{n=1}(-\hat{\Theta})^n\left(eE_i v^0_i+e^2 E_3B_3 \Omega_i v^0_i\right),\label{SSolution}
\eeqa
understanding that $\hat{\Theta}^n$ means that the operator $\hat{\Theta}$ is successively applied $n$ times. The main advantage of this method is that one can easily keep track of the desired power of magnetic field $eB_3$ ($\hat{\Theta}\sim O(eB_3)+O(e^2B^2_3)$ due to a $eB_3$ power in the function $g$). Also, this method is quite convenient when one consider the particular situation of $\bm{E}||\bm{B}$, that is
\beqa
f_1=-\frac{\tau}{D}\frac{\partial f_0}{\partial \varepsilon_{\bm{k}}}\left(eE_3 v^0_3+e^2 E_3B_3 \Omega_i v^0_i\right)-
\frac{\tau}{D}\frac{\partial f_0}{\partial \varepsilon_{\bm{k}}}\sum^{\infty}_{n=1}(-\hat{\Theta})^n\left(eE_3 v^0_3+e^2 E_3B_3 \Omega_i v^0_i\right),\label{Sfinalboltzmann}
\eeqa
because now, it can be seen explicitly that $\partial_\theta v^0_3=0$, and, after a (very) little bit of algebra, the product $\Omega_i\cdot v^0_i$ does not depend on $\theta$, $\partial_\theta (\Omega_i\cdot v^0_i)=0$. These two facts automatically kill all the second term of the RHS of (\ref{Sfinalboltzmann}) and we obtain the final solution of the Boltzmann equation
\beq
f_1=-\frac{\tau}{D}\frac{\partial f_0}{\partial \varepsilon_{\bm{k}}}\left( v^0_3+eB_3 \Omega_i v^0_i\right)eE_3.\label{Sfinalboltzmann2}
\eeq

The function $f_1$ in (\ref{Sfinalboltzmann2}) is the solution we will use from now on.

The third component of the current $J_3$ reads (after using (\ref{Svelocityeq}))
\beqa
J_{3}=-e^2E_3\tau\int (d\bm{k})\frac{\partial f_0}{\partial \varepsilon_{\bm{k}}}\left(\frac{1}{1+e\Omega_3 B_3}\right)\left( v^0_3+eB_3 \Omega_i v^0_i\right)\left(v^0_3+eB_3(\Omega_a\cdot v^0_a-\partial_3 m_3)-(eB_3)^2(\Omega_a\cdot \partial_a m_3)\right).\label{Smagnetocurrent}
\eeqa
From this equation, we can directly read the (magneto)conductivity $\sigma_{33}[B_3]$ as $J_3=\sigma_{33}E_3$. In the previous expression, we have substituted $f=f_0+f_1$ and noticed that the \emph{quasiparticle} anomalous Hall contribution 
\beq
J^H_i=e^2E_3\int (d\bm{k})f_0\epsilon_{i3j}\Omega_j,
\eeq
actually vanishes in the considered models because $\varepsilon_{\bm{k}}$ is an even function of $\bm{k}$ so does $f_0(\varepsilon_{\bm{k}})$, while the components $\Omega_{(1,2)}$ are odd functions of $\bm{k}$.
Let us note however, that this is not the \emph{topological} contribution to the Hall current that comes from the axial anomaly. This is the piece that comes from the quasiparticles at finite chemical potential. The contribution to the Hall current proportional to $f_1$ is actually proportional to $E^2_3$, and it goes beyond linear response and thus discarded here. It is important to say that this is true because we are dealing with the electromagnetic configuration $\bm{E}||\bm{B}$. For the configuration $\bm{B}\perp\bm{E}$, a non-vanishing contribution to the Hall conductivity will appear.

We also work at zero temperature, $T=0$ and finite chemical potential , so $\frac{\partial f_0}{\partial \varepsilon_{\bm{k}}}\sim -\delta\left(\mu-\varepsilon_{\bm{k}}\right)$, and we can safely work with one band. This is  why we computed $v^0_i$, $\Omega_i$, and $m_i$ for the conduction band and in terms of $\varepsilon_{\bm{k}}$. In all the expressions the delta function tells us to write $\mu$ instead of $\varepsilon_{\bm{k}}$ and to write $k$ as a function of $k_3$ and $\mu$ by the constraint $\mu=\varepsilon_{\bm{k}}$.

We have to keep in mind that $D$ also depends on $\bm{B}$, so it has to be expanded in powers of $eB_3$ as well: $\left(\frac{1}{1+e\Omega_3 B_3}\right)=1-(eB_3\Omega_3)+(eB_3\Omega_3)^2-...$
\subsection{Computation of the longitudinal magnetoconductivity}
As we said, expanding equation(\ref{Smagnetocurrent}) in powers of $eB_3$ up to second order, $\sigma_{33}[B_3]\simeq \sigma^{(0)}_{33}+\sigma^{(1)}_{33}(eB_3)+\sigma^{(2)}_{33}(eB_3)^2$ we have (after some algebraic simplifications):
\begin{subequations}
\beq
\sigma^{(0)}_{33}=-e^2\tau\int (d\bm{k})\frac{\partial f_0}{\partial \varepsilon_{\bm{k}}}\left(v^0_3\right)^2,\label{Scontri0}
\eeq
\beq
\sigma^{(1)}_{33}=-e^2\tau\int(d\bm{k})\frac{\partial f_0}{\partial \varepsilon_{\bm{k}}}\left(2\Omega_a v^0_a-\partial_3 m_3-v^0_3\Omega_3\right)v^0_3,\label{Scontri1}
\eeq
\beq
\sigma^{(2)}_{33}=-e^2\tau\int (d\bm{k})\frac{\partial f_0}{\partial \varepsilon_{\bm{k}}}\left((\Omega_a v^0_a-\Omega_3v^0_3)^2-(\Omega_av^0_a-v^0_3\Omega_3 )\partial_3 m_3-v^0_3\Omega_a\partial_a m_3\right).\label{Scontri2}
\eeq
\end{subequations}

Analytical expressions can be obtained for the conductivities $\sigma^{(0)}_{33}$ and $\sigma^{(1)}_{33}$ for all values of the Fermi level $\mu$ and $b_3$.
\subsection{Computation of $\sigma^{(0)}_{33}$}

In the case of $\sigma^{(0)}_{33}$, the computation is trivial. The $\delta$ function can be written as $\delta(\mu-\varepsilon_{\bm{k}})=\frac{\mu}{v^2k_+}\delta(k-k_+)$, with $k_+=\sqrt{\frac{\mu^2}{v^2}-\frac{b^2_3}{4}(1-\frac{k^2_3}{b^2_3})^2}$. With this and the expression for $v^0_3$, one can write, after writing $x=\frac{k_3}{b_3}$ and $\delta=\frac{2\mu}{vb_3}$:
\beq
\sigma^{(0)}_{33}=\frac{e^2\tau }{2^3\pi^2}\frac{vb^2_3}{\delta}\int dx x^2(1-x^2)^2\Theta[k_+],
\eeq
where $\Theta[k_{+}]=\Theta[\sqrt{\delta^2-(1-x^2)^2}]$ is the Heaviside step function of $k_+$ written in dimensionless units. The constraint imposed by the step function defines the limits of integration over the variable $x$. One has to distinguish two regions: $0<\delta<1$ and $\delta>1$. The reason is because the step function selects the part of the function $\sqrt{\delta^2-(1-x^2)^2}$ that is positive (in fact, because the square root, the function is positive whenever it exists). For $\delta<1$, we have to regions: $x\in[-\sqrt{1+\delta]},-\sqrt{1-\delta}]$, and $x\in[\sqrt{1-\delta]},\sqrt{1+\delta}]$ (physically corresponding to the regions around the Weyl nodes), while for $\delta>1$ where the notion of two valleys or nodes disappear, there is a single domain $x\in[-\sqrt{1+\delta},\sqrt{1+\delta}]$. We also note that the integrand and the integration domains are even under $x\rightarrow-x$ so, for $\delta\ll 1$,
\beq
\sigma^{(0)}_{33}=2\times \frac{e^2\tau}{8\pi^2}\frac{vb^2_3}{\delta}\int^{\sqrt{1+\delta}}_{\sqrt{1-\delta}} dx x^2(1-x^2)^2\simeq \frac{e^2\tau v}{3\pi^2 }\frac{\mu^2}{v^2},
\eeq
which is precisely the same result for the two unbounded Weyl fermions. Interestingly, one can go to the opposite regime and obtain, for $\delta\gg 1$
\beqa
\sigma^{(0)}_{33}=\frac{e^2\tau}{2^3\pi^2}\frac{v b^2_3}{\delta}\int^{\sqrt{1+\delta}}_{-\sqrt{1+\delta}} dx x^2(1-x^2)^2\simeq
\frac{e^2\tau v}{7\pi^2}\frac{\mu^2}{v^2}\left(\frac{2\mu}{vb_3}\right)^\frac{1}{2}.
\eeqa

\subsubsection{Computation of $\sigma^{(1)}_{33}$}
The computation of $\sigma^{(1)}_{33}$ requires is more time-consuming since the algebra is more tedious (but straightforward in any case). For any value of $\mu$, $\sigma^{(1)}_{33}$ is always positive and a monotonously increasing function of $\mu$. Since $v^0_3$ is an odd function of $k_3$ while $\Omega_i$ is even and $\partial_3 m_3$ is odd, the expression for $\mu$, $\sigma^{(1)}_{33}$ can be further simplified to 
\beq
\sigma^{(1)}_{33}=-e^3B_3\tau\int(d\bm{k})\frac{\partial f_0}{\partial \varepsilon_{\bm{k}}}\left(\Omega_3 v^0_3-\partial_3 m_3\right)v^0_3.\label{Scontri11}
\eeq

The analytical expression becomes more complex:
\beqa
\sigma^{(1)}_{33}=\frac{e^3\tau B_3}{4\pi^2}\frac{v}{\delta^4}\int dxx^2(1-x^2)\cdot
\left(x^6+(1-2x^2)(\delta^2+x^2)\right)\Theta[k_+].
\eeqa 

Performing the integral in the limit $\delta\ll 1$, one gets

\beq
\sigma^{(1)}_{33}\simeq\frac{8}{15}\frac{e^2\tau v}{\pi^2}\frac{\mu}{v b_3}.
\eeq

In the opposite regime, $\delta\gg1$,
\beq
\sigma^{(1)}_{33}\simeq\frac{15}{77}\frac{e^2\tau v}{\pi^2}\frac{\mu}{v b_3}\left(\frac{2\mu}{vb_3}\right)^{\frac{1}{2}}.
\eeq

\subsection{Computation of $\sigma^{(2)}_{33}$}

With $\sigma^{(0)}_{33}$ and $\sigma^{(1)}_{33}$, it was possible to perform a brute-force calculation. It is out of question to do the same for $\sigma^{(2)}_{33}$ from scratch. What we can do is to compute separately each part and sum them all up at the end.

As it is written, we can compute separately the three pieces in (\ref{Scontri2}). After a little bit of algebra, it can be seen that the second term in (\ref{Scontri2}) is actually an odd function of $k_3$, so, being always the integration contour even in $k_3$, we can safely say that this part is identically zero, leaving only with the first and third terms in (\ref{Scontri2}).
Again, after some algebra, the computation can be done analytically. 

Performing similar algebraic manipulations as before, the expression for $\sigma^{(2)}_{33}$ is
\beqa
\sigma^{(2)}_{33}=\frac{e^4\tau v^3 B^2_3}{2\pi^2b^2_{3}\delta^{7}}\int dx (\delta^4-(1-x^2)^4(2x^6-1)-
(1-x^2)^2(2x^6-x^4-2)\delta^2)\Theta[k_+].
\eeqa
Performing the integral and expanding in powers of $\delta$ for the case of $\delta\ll1$, we get
\beq
\sigma^{(2)}_{33}\simeq\frac{e^4\tau v}{2\pi^2}\left(\frac{7}{30}\frac{v^2}{\mu^2}+\frac{47}{140}\frac{1}{b^2_3}\right)B^2_3
\eeq
In the opposite limit, $\delta\gg1$,
\beq
\sigma^{(2)}_{33}\simeq \frac{16}{165}\frac{e^4\tau v}{\pi^2}\frac{1}{b^2_3}\left(\frac{2\mu}{v b_3}\right)^{\frac{1}{2}}B^2_3.
\eeq
\subsection{Computation of the transverse magnetoconductivity}
Let us compute the transverse magnetoconductivity. In this case, without loss of generality, we will consider $\bm{E}=E_1\hat{\bm{x}}$ keeping the magnetic field pointing along the third direction ($\bm{E}\perp\bm{B}$). The formal solution of the Boltzmann equation (\ref{SSolution}) is now,
\beqa
f_1=-\frac{\tau}{D}\frac{\partial f_0}{\partial \varepsilon_{\bm{k}}}\left(eE_1 v^0_1\right)-
eE_1\frac{\tau}{D}\frac{\partial f_0}{\partial \varepsilon_{\bm{k}}} \sum^{\infty}_{n=1}(-\hat{\Theta})^n[v^0_1],\label{STranssol}
\eeqa
where the application of the operator $\hat{\Theta}$ on $v^0_1$ does not vanish, because $v^0_1=v^2k_1/\varepsilon_{\bm{k}}$ explicitly depends on the angle $\theta$.

Using (\ref{Svelocityeq}) and (\ref{STranssol}) up to first order in powers of $eB_3$, leads to the following expressions for the transverse magnetoconductivity:
\begin{subequations}
\beq
\sigma^{(0)}_{11}=-e^2\tau\int \frac{d^3\bm{k}}{(2\pi)^3}\frac{\partial f_{0}}{\partial\varepsilon}\left(v^0_1\right)^2,\label{SLMC0}
\eeq
\beqa
\sigma^{(1)}_{11}&=&e^3\tau B_3\int \frac{d^3\bm{k}}{(2\pi)^3}\frac{\partial f_{0}}{\partial\varepsilon}(v^0_1\partial_1m_3+(v^0_1)^2\Omega_3+\\\nonumber
&+&\tau\frac{v^2}{\varepsilon_{\bm{k}}}v^0_1\partial_{\theta}v^0_1).\label{SLMC1}
\eeqa
\end{subequations}
It can be seen that the last term in the right hand side of (\ref{SLMC1}) vanishes upon angular integration. It is important to note that the contribution linear in $B_3$ has opposite sign than the zeroth component.

As in the previous cases, we can compute analytically the contributions to the transverse magnetoconductivity. 
The zeroth contribution to the conductivity is
\beq
\sigma^{(0)}_{11}=\frac{e^2 v\tau}{16\pi^2}\frac{b^2_3}{\delta}\int dx (\delta^2-(1-x^2)^2)\Theta[k_+],
\eeq 
or, in the limit $\mu\ll vb_3$,
\beq
\sigma^{(0)}_{11}\simeq \frac{e^2 \tau v}{3\pi^2}\frac{\mu^2}{v^2}.
\eeq

The contribution linear in the magnetic field is, after writing all que variables in their dimensionless form,
\beq
\sigma^{(1)}_{11}=\frac{3}{8}\frac{e^3 v\tau B_3}{\pi^2}\frac{1}{\delta^4}\int dx x^2(1-x^2)(\delta^2-(1-x^2)^2). 
\eeq
As before, performing the integral and expanding in powers of $\mu \ll vb_3$, we obtain

\beq
\sigma^{(1)}_{11}\simeq -\frac{1}{10}\frac{e^3 v \tau B_3}{\pi^2}\frac{\mu}{vb_3}.
\eeq
\subsection{Explicit demonstration of the dependence of $\bm{b}\cdot\bm{B}$ of the longitudinal magnetoconductivity}
In the previous sections, we have computed the magnetoconductivity using $\bm{B}||\bm{b}$. In this section we will demonstrate that the magnetochiral effect enters as the scalar product $\bm{B}\cdot\bm{b}$ by showing that, when the magnetic field is perpendicular to $\bm{b}$, and to linear order in the magnetic field, the magnetochiral effect is zero. We will do it in the case of the longitudinal magnetoconductivity, that is, with $\bm{E}||\bm{B}$. Without loss of generality, we will choose $\bm{B}=B_2 \hat{\bm{y}}$, and $\bm{E}=E_2\hat{\bm{y}}$.

Let us write again the Boltzmann equation:
\beq
f_1+\frac{\tau}{D}D\dot{\bm{k}}\cdot\partial_{\bm{k}}f_1=-\frac{\tau}{D}D\dot{\bm{k}}\cdot\bm{v}_{0}\frac{\partial f_0}{\partial \varepsilon}.
\eeq
With the chosen electromagnetic configuration, it is easy to see that
\beq
D\dot{\bm{k}}\cdot\bm{v}_{0}=eE_2 v^0_2+e^2E_2 B_2(\Omega_av^0_a)+e^2B^2_2(v^0_1\partial_3 m_2-v^0_3\partial_1 m_2),
\eeq
and
\beq
D\dot{\bm{k}}\cdot\partial_{\bm{k}}=eB_2(v_1\partial_3-v_3\partial_1).
\eeq
Also, in this case, $D=1+e\bm{\Omega}\cdot\bm{B}=1+eB_2\Omega_2$.

As before, using the Zener-Jones method, we can formally write the solution of this Boltzmann equation as
\beq
f_1=-\tau\frac{1}{1+\hat{\Theta}}\left[\frac{\partial f_0}{\partial \varepsilon}\frac{1}{D}D\dot{\bm{k}}\cdot\bm{v}_{0}\right]=-\tau(1+\sum^{\infty}_{n=1}(-\hat{\Theta})^n)\left[\frac{\partial f_0}{\partial \varepsilon}\frac{1}{D}D\dot{\bm{k}}\cdot\bm{v}_{0}\right],
\eeq
being in this case $\hat{\Theta}=\frac{\tau}{D}eB_2(v_1\partial_3-v_3\partial_1)$.

Since we are interested in the fate of the magnetochiral effect, we will content ourselves with retaining in the computation only terms up to first order in $B_2$. The operator $\hat{\Theta}$ is already first order in $B_2$ so, 
\beq
f_1=-\frac{\tau}{D}\frac{\partial f_0}{\partial \varepsilon}D\dot{\bm{k}}\cdot\bm{v}_{0}+\tau\frac{\partial f_0}{\partial \varepsilon}\hat{\Theta}\left[\frac{1}{D}D\dot{\bm{k}}\cdot\bm{v}_{0}\right]+\tau\frac{1}{D}D\dot{\bm{k}}\cdot\bm{v}_{0}\hat{\Theta}\left[\frac{\partial f_0}{\partial \varepsilon}\right],
\eeq
or
\beq
f_1=-\frac{\tau}{D}\frac{\partial f_0}{\partial \varepsilon}D\dot{\bm{k}}\cdot\bm{v}_{0}+\tau\frac{\partial f_0}{\partial \varepsilon}D\dot{\bm{k}}\cdot\bm{v}_{0}\hat{\Theta}\left[\frac{1}{D}\right]+\frac{\tau}{D}\frac{\partial f_0}{\partial \varepsilon}\hat{\Theta}\left[D\dot{\bm{k}}\cdot\bm{v}_{0}\right]+\tau\frac{1}{D}D\dot{\bm{k}}\cdot\bm{v}_{0}\hat{\Theta}\left[\frac{\partial f_0}{\partial \varepsilon}\right].
\eeq
The last term in the previous equations is zero because $\hat{\Theta}\left[\frac{\partial f_0}{\partial \varepsilon}\right]=eB_2(v^0_1\frac{\partial}{\partial k_3}\frac{\partial f_0}{\partial \varepsilon}-v^0_3\frac{\partial}{\partial k_1}\frac{\partial f_0}{\partial \varepsilon})=eB_2(v^0_1v^0_3-v^0_3v^0_1)\frac{\partial^2 f_0}{\partial \varepsilon^2}=0$.

As we said, $\hat{\Theta}\sim O(B_2)$ so $\hat{\Theta}[\frac{1}{D}]\sim O(B^2_2)$, and we neglect the second term of the r.h.s. of the previous equation. Also, to this order in $B_1$, we can set $D\sim 1$ in the third term and $1/D\simeq 1-eB_2\Omega_2$ in the first term:
\beq
f_1\simeq-\tau (1-eB_2\Omega_2)\frac{\partial f_0}{\partial \varepsilon}D\dot{\bm{k}}\cdot\bm{v}_{0}+\tau\frac{\partial f_0}{\partial \varepsilon}\hat{\Theta}\left[D\dot{\bm{k}}\cdot\bm{v}_{0}\right].
\eeq

Let us work out the term $\hat{\Theta}\left[D\dot{\bm{k}}\cdot\bm{v}_{0}\right]$ noticing that the velocity $\bm{v}$ is $\bm{v}=\bm{v}_0-eB_2\partial_{\bm{k}}m_2$:
\beqa
\hat{\Theta}\left[D\dot{\bm{k}}\cdot\bm{v}_{0}\right]= \tau eB_2(v_1\partial_3-v_3\partial_1)\left[D\dot{\bm{k}}\cdot\bm{v}_{0}\right]\simeq \tau eB_2(v^0_1\partial_3-v^0_3\partial_1)\left[D\dot{\bm{k}}\cdot\bm{v}_{0}\right].
\eeqa
To first order in $B_2$, we have $D\dot{\bm{k}}\cdot\bm{v}_{0}\simeq eE_2 v^0_2+e^2E_2 B_2(\Omega_av^0_a)$, so
\beq
\tau eB_2(v^0_1\partial_3-v^0_3\partial_1)\left[D\dot{\bm{k}}\cdot\bm{v}_{0}\right]=\tau e^2E_2B_2(v^0_1\partial_3-v^0_3\partial_1)\left[v^0_2\right].
\eeq
Putting all together, we finally have
\beqa
f_1&\simeq&-\tau (1-eB_2\Omega_2)\frac{\partial f_0}{\partial \varepsilon}(eE_2 v^0_2+e^2E_2 B_2(\Omega_av^0_a))+\tau^2e^2E_2B_2(v^0_1\partial_3-v^0_3\partial_1)\left[v^0_2\right]\simeq\nonumber\\&\simeq&-\tau\frac{\partial f_0}{\partial \varepsilon}eE_2 v^0_2-\tau\frac{\partial f_0}{\partial \varepsilon}e^2E_2 B_2(\Omega_av^0_a-\Omega_2v^0_2)+\tau^2e^2E_2B_2(v^0_1\partial_3v^0_2-v^0_3\partial_1v^0_2).
\eeqa

Now, we have to use this expression for $f_1$ in the definition of the quasiparticle current $J_2=e\int (d\bm{k})f_1D\dot{x}_2$. Making use of the expression for $D\dot{\bm{x}}$, we have, to first order in $B_2$, $D\dot{x}_2\simeq v^0_2-eB_2\partial_2 m_2+eB_2(\Omega_a v^{0}_a)$, so, we obtain the following expression for $J_2$ up to first order in $B_2$:

\beqa
J_2&=&-e^2E_2\tau\int (d\bm{k})\frac{\partial f_0}{\partial \varepsilon}(v^0_2)^2-\nonumber\\
&-&e^3E_2B_2\tau\int (d\bm{k})\frac{\partial f_0}{\partial \varepsilon}\left(2\Omega_av^0_av^0_2-\Omega_2(v^0_2)^2-\partial_2 m_2 v^0_2\right)+\nonumber\\
&+&e^3E_2B_2\tau^2\int (d\bm{k})\frac{\partial f_0}{\partial \varepsilon}\left(v^0_1\partial_3v^0_2-v^0_3\partial_1v^0_2\right).
\eeqa

As before, the first term defines the longitudinal conductivity in absence of magnetic field: $\sigma^{(0)}_{22}=-e^2\tau \int (d\bm{k})\frac{\partial f_0}{\partial \varepsilon}(v^0_2)^2$.
In the case of being non-zero, the next terms would define the magnetochiral component of the magnetoconductivity:

\beqa
\sigma^{(1)}_{22}=-e^3B_2\tau\int (d\bm{k})\frac{\partial f_0}{\partial \varepsilon}\left(2\Omega_av^0_av^0_2-\Omega_2(v^0_2)^2-\partial_2 m_2 v^0_2-\tau (v^0_1\partial_3v^0_2-v^0_3\partial_1v^0_2)\right).
\eeqa

We can compute each term of the integrand of the previous expression using that, at $T=0$, $\frac{\partial f_0}{\partial \varepsilon}\simeq -\delta(\mu-\varepsilon_{\bm{k}})$, so the integral in momentum can be written as $\int (d\bm{k})\delta(\mu-\varepsilon_{\bm{k}})\rightarrow\frac{1}{8\pi^3}\int dk_3\int d\theta\int dk k\frac{\mu}{v^2k_{+}}\delta(k-k_+)$, with $k_+=\sqrt{\frac{\mu^2}{v^2}-\frac{b^2_3}{4}(1-\frac{k^2_3}{b^2_3})^2}$. Substituting $\varepsilon_{\bm{k}}$ by $\mu$ and $k$ by $k_+$, we get, after some straightforward but tedious algebra:
\begin{subequations}
\beq
\Omega_a v^0_a v^0_2=v^7\frac{(k^7_3-2k^5_3b^2_3+k^4_3b^3_3+k^3_3b^4_3-2k^2_3b^5_3+b^7_3-\frac{4\mu^2b^5_3}{v^2})(\frac{\mu^2}{v^2}-\frac{(k^2_3-b^2_3)^2}{4b^2_3})^{\frac{1}{2}}}{8\mu^5 b^5_3}\sin{\theta},
\eeq
\beq
\Omega_2(v^0_2)^2=-v^7\frac{(\frac{\mu^2}{v^2}-\frac{(k^2_3-b^2_3)^2}{4b^2_3})^{\frac{3}{2}}}{16\mu^5}\sin{\theta},
\eeq
\beq
\partial_2m_2 v^0_2=v^7\frac{((b^2_3-k^2_3)^2\sin^2{\theta}-\frac{2\mu^2}{v^2}b^2_3)(\frac{\mu^2}{v^2}-\frac{(k^2_3-b^2_3)^2}{4b^2_3})^{\frac{1}{2}}}{4\mu^5 b^2_3}\sin{\theta}
\eeq
\beq
v^0_1\partial_3v^0_2-v^0_3\partial_1v^0_2=0(!).
\eeq
\end{subequations}
The last expression turns out to be identically zero. 

Now, it is easy to see that, despite the rather involved dependence with the momentum $k_3$, all the expressions are composed by terms that are odd powers of $\sin{\theta}$ and all give zero after angular integration.

Thus, all this implies that 

\beq
\boxed{\sigma^{(1)}_{22}=0.}
\eeq
and there is no magnetochiral effect when $\bm{B}\perp \bm{b}$, or, equivalently, the magnetochiral effect enters as the scalar product $\bm{B}\cdot\bm{b}$.
\end{widetext}
\end{document}